\documentstyle[preprint,prb,aps]{revtex}
\begin{document}
\newcommand{\beq}{\begin{equation}}    
\newcommand{\eeq}{\end{equation}}      
\draft
\title{ \bf Finite Temperature Resonant Magnetotunneling in
AlGaAs-GaAs-AlGaAs Heterostructures}
\author{
{\O}. Lund B{\o},$^{(1)}$ Yu. Galperin,$^{(1,2)}$
and K.~A.~Chao,$^{(3)}$}
\address{
$^{(1)}$Department of Physics, University of Oslo,  P. O. Box 1048
Blindern, N 0316 Oslo 3, Norway,}
\address{$^{(2)}$ A. F. Ioffe
Physico-Technical  Institute, 194021 St. Petersburg, Russia,}
\address{
$^{(3)}$Department of Physics, Norwegian Institute of Technology,
The University of  Trondheim, N~7034 Trondheim, Norway.}

\date{\today}
\maketitle
\begin{abstract}
We have analyzed the effect of electron-LO phonon interaction in a
double-barrier resonant tunneling structure under a magnetic field
{\bf B} applied parallel to the tunneling current. While the low
temperature anti-crossing phenomenon has already been investigated,
here we study the phonon absorption resonant magnetotunneling at
finite temperatures. The Matsubara technique is used to sum up
resonant diagrams of higher orders, and the result is
selfconsistently renormalized. The phonon absorption tunneling
spectrum has been calculated numerically.
The result shows that the phonon-absorption process produces
two broad inelastic wings on the main elastic tunneling peak,
the strength of which is sensitive to
the resonant condition, and grows with increasing temperature. The
width of the inelastic wings are proportional to $B^{1/2}$. This is
in contrast to the appearance of anti-crossing in the phonon
emission tunneling process. The difference is due to the fact that
the phonon emission is a coherent process, while the absorption is a
typical incoherent process since absorbed real phonons have random
phases.
\end{abstract}
\pacs{ 71.38.+i, 72.10.Di, 73.40.Kp }

\narrowtext
\section{Introduction}

Since the pioneer work of Tsu and Esaki \cite{esaki} and the
observation of negative differential resistance by Sollner {\em et
al.} \cite{sollner}, there has been increasing interest in resonant
tunneling through a double-barrier resonant tunneling structure
(DBRTS) both experimentally and theoretically. Among the important
tools of characterizing DBRTS, one is the DC transport
\cite{azbel,price,luryi,weil,frensley} together with phonon induced
or laser induced tunneling
\cite{goldman,glazman,wilk1,wilk,cai,sokolovski,jons,johnsson}.
In connection to such characterization, the DC current noise of a
DBRTS has also been investigated. A common feature of various resonant
tunneling structures is the {\em inelastic} resonant tunneling due to
the coupling of tunneling electrons to phonons. This phenomenon is not
only interesting but also useful because it provides a way to
investigate both electron modes and phonon modes. A theory of phonon
assisted resonant tunneling through a DBRTS in the absence of an
external magnetic field was presented by Wingreen {\em et al.}
\cite{wilk} and by Jonson \cite{jons}. Assuming a perfect sample
without imperfection and impurity, in these works an exact solution
was obtained when the system is reduced to one-dimension and when
tunneling processes through different quasibound states in the well
are decoupled. The problem becomes much complicated when these
tunneling processes are coupled, for example by phonons.

Magnetic field is another important tool for sample characterization
because of the formation of Landau level spectrum and the drastic
modification of electronic wave functions, especially under strong
external field. For a DBRTS, the challenging situation is when the
magnetic {\bf B} is applied parallel to the tunneling current {\bf
I}, as schematically illustrated in Fig.~1. To clarify the important
features of Fig.~1, let us consider the experimentally most
extensively studied DBRTS
$GaAs^+/Al_{0.3}Ga_{0.7}As/GaAs/Al_{0.3}Ga_{0.7}As/GaAs^+$, with both
the barrier width and the well width of the order of 40-60 {\AA}. In
this case the barrier height is about 300 $meV$, and in the well
there exists only one quasibound state with energy $\epsilon_b$. We
will set the zero reference energy at the bottom of the conduction
band of the collector (labeled by $c$), and so the conduction band
minimum of the emitter (labeled by $e$) is raised to $eV$ by the
bias $V$. Under a magnetic field {\bf B} perpendicular to interfaces
which are parallel to the $xy$-plane, the quasibound energy level
$\epsilon_b$ splits into a series of discrete Landau levels
$\epsilon_b+(n+\frac{1}{2})\hbar\omega_c$, where $\omega_c=eB/m^*c$
is the cyclotron frequency in $GaAs$. The electronic energy levels in
the emitter and collector, on the other hand, form sets of Landau
bands $eV+\epsilon_z(k_{e,z})+(n_e+\frac{1}{2})\hbar\omega_c$
and $\epsilon_z(k_{c,z})+(n_c+\frac{1}{2})\hbar\omega_c$,
respectively, where $\epsilon_z(k_z)=\hbar^2k^2_z/2m^*$ is the kinetic
energy along the tunneling direction.

Such level splitting has important influences on both elastic and
phonon assisted tunneling \cite{mendes,eaves,yang}. In a perfect
DBRTS, if the electron-phonon interaction is ignored the tunneling
tunneling process conserves the Landau level quantum number
($n_e=n=n_c$). If the magnetic field is sufficiently strong such
that $\hbar\omega_c$ is much larger than the thermal energy
$k_{\text B}T$, the resonant tunneling is indeed one-dimensional with
all tunneling channels decoupled from each other. The electron-phonon
interaction mixes the channels. If the magnetic field is tuned into
the resonant condition $\hbar\omega_c=\hbar\omega_0$, where
$\omega_0$ is the optical phonon frequency in $GaAs$, then the
resonant tunneling processes through two adjacent Landau levels are
resonantly coupled. In fact, this is one way to observe the strong
renormalization of Landau levels due to the electron-phonon
interaction \cite{sarma1,boeb}. In a 3D electron gas, the
renormalization effect is proportional to $g^{4/3}$, while in a 2D
system it is proportional to $g$, where $g$ is the electron-phonon
coupling constant to be defined later.

Under a strong magnetic field and at low temperatures, one relevant
interesting phenomenon is the formation of magnetopolaron. Let us
consider the simpler situation that only the lowest Landau band in
the emitter ($n_e=0$) is occupied with quasi Fermi energy
$\epsilon_{e,\text{F}}$. In a perfect DBRTS, one might expect to see
in the $I$-$V$ characteristics the elastic resonant tunneling peak
around the bias $eV=\epsilon_b-\epsilon_{e,\text{F}}$, with series of
phonon replicas at
$eV=\epsilon_b+n\hbar\omega_c-\epsilon_{e,\text{F}}+\nu\hbar\omega_0$,
where $\nu$ is an integer. However, complication arises when the
condition $\omega_c=(\nu-\nu ')\omega_0/(n'-n)$ is satisfied. Then, an
electron in the emitter can tunnel resonantly into two coherent
degenerate states in the well: into the Landau level $n$ by emitting
$\nu$ phonons, or into the Laudau level $n'$ by emitting $\nu '$
phonons. The two final states are strongly hybridized because of the
coherence, and so in phonon emission resonant magnetotunneling
experiments, magnetopolarons have been observed \cite{boeb,cyywi}. A
standard way to present such experimental data is to plot resonant
tunnelings peak position as a function of the magnetic field and the
bias voltage. In this two-dimensional plot, the existence of
magnetopolarons is reflected by the {\em anti-crossing}.

A phonon emission resonant magnetotunneling process leading to two
coherent final states, indicated as $E0$ and $E1$, is illustrated in
part (a) of Fig.~2, where the thick lines form the Landau fan in the
emitter, and the thin lines form the Landau fan in the well. Under
this resonant condition, the $I$-$V$ characteristics will exhibit two
resolved peaks of almost equal strength \cite{boeb,zcg}. A similar
diagram for phonon absorption resonant magnetotunneling is
demonstrated in part (b) of Fig.~2. Here again we assume that only
the lowest Landau level in the emitter is occupied, and the bias is
tuned into elastic resonant tunneling
$eV=\varepsilon_b-\varepsilon_{e,\text{F}}$. Then, the two sets of
Landau fans coincide with each other. At the magnetic field strength
$\hbar\omega_c=\hbar\omega_0$, an electron in the emitter having the
proper energy can tunnel resonantly into the state $A0$ elastically,
or into the state $A1$ by absorbing a phonon. We are interested in
finding out how the energy levels $A0$ and $A1$ would be modified when
they are resonantly coupled by phonons. In other words, we would like
to derive the resulting transmission spectrum when the phonon
absorption resonant magnetotunneling takes place.

The essential issue here is the coherence. Since absorbed thermal
phonons have random phases, they can not participate in coherent
process. This problem was investigated in our recent work \cite{zcg},
where relevant references are listed. However, this work
is based on a low order perturbation expansion, together with an
exponential renormalization (see e. g. Ref.~\onlinecite{mahan}). In
the vicinity of resonance, low
order perturbation expansion is not sufficiently accurate. In the
present work we improve our previous results with an infinite
summation of resonant diagrams and a selfconsistent renormalization.
Analytical expressions can be derived under the reasonable assumption
of a dispersionless longitudinal optical (LO) phonon branch. To
present a complete analysis, we will treat both the phonon emission
and the phonon absorption resonant magnetotunneling.

Sec.~II describes our model Hamiltonian, from which the transmission
probability and the differential conductance will be derived in
Sec.~III. The self-energy and the vertex correction will first be
analyzed at low order for the general case, and then be summed up to
all orders for a model two-level system, which contains the essential
physics for phonon assisted resonant magnetotunneling through a DBRTS.
This will be done in Sec.~VI, and the so-obtained self-energy is
renormalized selfconsistently in Sec.~V. Numerical results will be
presented in Sec.~VI, followed by some remarks in the last Sec.~VII.
The electron-phonon scattering matrix elements are calculated in
Appendix A.

\section{The model}

To describe our model system, let us refer to Fig.~1 which specifies
the potential and various energy levels of a DBRTS under a bias $V$,
with zero reference energy set at the conduction band minimum of the
collector. We assume the same material for the emitter, the well, and
the collector, and also assume only one quasibound level $\epsilon_b$
in the well, which is usually the case for realistic samples. In our
model we will ignore the scattering due to the interface roughness
and impurities. Since the magnetic field {\bf B} is applied in the
positive direction of $z$-axis perpendicular to interfaces, it is
convenient to choose the Landau gauge with the vector potential
{\bf A}=$(0,Bx,0)$.

The eigenstates in the well are labeled by the Landau level quantum
number $n$ and the index of degeneracy $k_y$. However, in the emitter
(or collector), besides the similar quantum numbers $(n_e,k_{e,y})$
[or $(n_c,k_{c,y})$], we need also to specify the kinetic energy
$\epsilon_z(k_{e,z})$ [or $\epsilon_z(k_{c,z})]$ of an electron along
the tunneling direction. For the convenience of presenting mathematical
formulas, we define the simplified notation $\alpha\equiv (n,k_y)$ for
the states in the well, and $\beta\equiv (n_j,k_{j,y},k_{j,z})$, where
$j$=$e$ for the emitter, and $j$=$c$ for the collector. In terms of
these notations, the normalized eigenstates
and the eigenvalues are
\begin{eqnarray}
\phi_{\alpha}({\bf r})& =& \chi(z)\exp (ik_yy) \varphi_n(x+l^2k_y) \, ,
 \\
E_{\alpha}& =& \epsilon_b + \varepsilon_n
           \equiv \epsilon_b + \hbar\omega_c (n+\frac{1}{2})
\end{eqnarray}
in the well, and
\begin{eqnarray}
\phi_{j,\beta}({\bf r})& =& \exp (ik_{j,z}z+ik_{j,y}y)
\varphi_n(x+l^2k_{j,y}) \, ,
\\
E_{j,\beta} &=& eV\delta_{j,e} + \epsilon_z(k_{j,z}) + \varepsilon_{n_j}
\nonumber \\
            &\equiv&  eV\delta_{j,e} + \epsilon_z(k_{j,z})
                   + \hbar\omega_c (n_j+\frac{1}{2})
\end{eqnarray}
in the emitter ($j=e$) under the bias $eV$, and the collector
($j=c$). Here $l=\sqrt{c\hbar/eB}$ is the magnetic length.

Using this set of basis functions, our model system in Fig.~1 can be
described with the tunneling Hamiltonian
\beq
{\cal H} = {\cal H}_{\text e} + {\cal H}_{\text {ph}} + {\cal
H}_{\text{e-ph} }\, .
 \eeq
The electronic part of the Hamiltonian with an external magnetic field
is given by
\begin{eqnarray}
{\cal H}_{\text e} &=& \sum_{j,\beta}
E_{j,\beta} c_{j,\beta}^\dagger c_{j,\beta}
+ \sum_\alpha E_\alpha c_\alpha^\dagger c_\alpha \nonumber \\
& & + \sum_{j,\alpha,\beta}
\left[ V_{j,\beta\alpha}c_\alpha^\dagger c_{j,\beta}+h.c\right] \, .
\end{eqnarray}
The tunnel matrix elements $V_{j,\beta\alpha}$ have to  be calculated using
the eigenstates listed above. Since the interfaces are assumed to
be perfect, the momentum $\bf k_{\|}$ parallel to interfaces remains
invariant during the tunneling through a potential barrier. Both the
Landau level index and $k_y$ are then conserved, and so the calculation
of the matrix elements $V_{j,\beta\alpha}$ reduces to the solving of a
one-dimensional Schr\"{o}dinger equation\cite{zrc}, following with the
application of the Bardeen's prescription\cite{bar}.

Electrons interact with LO phonons. Here we will neglect the weak
phonon dispersion, and the resulting phonon Hamiltonian is simply
\beq
{\cal H}_{\text{ph}}=\hbar\omega_0\sum_{\bf q}
b_{\bf q}^{\dagger}b_{\bf q} \, .
\eeq
Since electrons in the well has a finite life time, it is reasonable
to expect that the dominating contribution to electron-LO phonon
interaction comes from the situation when electrons occupy the
quasibound state in the well \cite{jons}. In terms of the Fr\"{o}lich
Hamiltonian\cite{mahan}, the electron-LO phonon interaction can be
expressed as
\beq
{\cal H}_{\text{e-ph}}=\sum_{\alpha ,\alpha_1,{\bf q}}
M_{\alpha\alpha_1}({\bf q})(b_{\bf q}^\dagger+b_{\bf q})
c_{\alpha_1}^\dagger c_{\alpha} \, ,
\eeq
where $\alpha =(n,k_y)$, $\alpha_1=(n_1,k_y-q_y)$, and
\begin{eqnarray}\label{matr}
M_{\alpha\alpha_1}({\bf q}) &\equiv& \frac{M}{\sqrt{V_0}q}
\int \phi^*_{\alpha_1}({\bf r}) e^{i{\bf q}\cdot{\bf r}}
\phi_{\alpha}({\bf r}) d{\bf r} \, , \nonumber \\
M^2 &=& \pi e^2 \hbar\omega_0\left( \frac{1}{\epsilon_\infty}
-\frac{1}{\epsilon_0} \right) \, .
\end{eqnarray}

The above Hamiltonian describes the general case. We are interested in
the phonon absorption resonant magnetotunneling process as illustrated
by part (b) in Fig.~2. Since the cross section for multiphonon
absorption is much smaller than that for single phonon absorption, our
problem can be well represented by retaining only the two lowest Landau
levels $n=0$ and $n=1$. With low impurity concentration in the emitter
such that only the lowest Landau level $n_e=0$ in the emitter is
occupied, the resulting two-level system can be solved exactly. This
exact solution can be extended to study the phonon emission resonant
magnetotunneling if the process demonstrated by part (a) in Fig.~2 will
be modified as follows. We first increase the impurity concentration in
the emitter such that under very strong magnetic field, only the two
lowest Landau levels $n_e=0$ and $n_e=1$ are occupied. Next we set the bias
at $eV=\epsilon_b+\hbar\omega_0-\epsilon_{e,\text{F}}$. Then, when the
magnetic field is tuned to $\hbar\omega_c=\hbar\omega_0$, electrons in
the second Landau level ($n_e=1$) in the emitter can tunnel resonantly
either into the $E1$ state via an elastic process, and into the $E0$
state by emitting a phonon. This is again a resonant tunneling in a
two-level system with two coherent degenerate final states.

In the rest of this paper, we will first perform a general theoretical
analysis. Then we will continue to study in details this exactly
solvable two-level system.

\section{Transmission probability}

To calculate the conductance of a DBRTS, we must first calculate the
transmission probability of an electron from the initial state
$(n_e,k_{e,y},k_{e,z})$ in the emitter with energy $\varepsilon$ to the
final state $(n_c,k_{c,y},k_{c,z})$ in the collector with energy
$\varepsilon_1$. This transmission probability can be expressed in terms
of the Fourier transform of the two-particle Green's function as
\cite{wilk}
\begin{eqnarray}
& & T_{\alpha\alpha_1}(\varepsilon,\varepsilon_1) =
\gamma_e(\varepsilon)\gamma_c(\varepsilon_1) \nonumber \\
& & \times\int {d\tau \, ds \, dt \over 2 \pi \hbar^3}
e^{i[(\varepsilon -\varepsilon_1)\tau +\varepsilon_1 t -
\varepsilon s]/\hbar} K_{\alpha\alpha_1}(\tau,s,t) \, ,
\end{eqnarray}
where
\begin{eqnarray}
K_{\alpha\alpha_1}(\tau,s,t) &=& \Theta(s) \Theta(t) \nonumber \\
& & \times \left<
c_{\alpha}(\tau -s)c^{\dag}_{\alpha_1}(\tau)
c_{\alpha_1}(t)c^{\dag}_{\alpha}(0)\right>
\end{eqnarray}
is the so-called {\em transmission Green's function}. Here
$\alpha =(n,k_y)\equiv (n_e,k_{e,y})$ and
$\alpha_1=(n_1,k_{1y})\equiv (n_c,k_{c,y})$ label two electronic states
in the well. \mbox{$\gamma_{j}(\varepsilon)=2\pi\sum_\beta
|V_{j,\beta\alpha}|^2\delta (\varepsilon -E_{j,\beta})$}
is the escape rate of an electron from the well to the emitter $j=e$
or the collector $j=c$. The total escape rate from the well is simply
$\gamma (\varepsilon)=\gamma_e(\varepsilon)+\gamma_c(\varepsilon)$.

It can been shown \cite{wilk,jons,sarma} that the Fourier transform
$K_{\alpha\alpha_1}(\varepsilon,\varepsilon_1)$ of
$K_{\alpha\alpha_1}(\tau,s,t)$ is in fact equivalent to a two-particle Green's
function with input energies $(\varepsilon,\varepsilon_1-\Omega)$
and output energies $(\varepsilon_1,\varepsilon -\Omega)$, analytically
continued into the upper half plane of the complex variables
$(\varepsilon,\varepsilon_1)$ and into the lower half plane of the
complex variables $(\varepsilon -\Omega,\varepsilon_1-\Omega)$. This
two-particle Green's function is schematically illustrated in Fig.~3.
The diagrammatic technique for an interacting electron-phonon system
under a uniform external magnetic field has been extensively discussed
in the literature \cite{lev1,lev2,lms}, and here we will follow the
conventional approach. After completing such calculation and then
setting $\Omega\rightarrow 0$, we obtain
$K_{\alpha\alpha_1}(\varepsilon,\varepsilon_1)$ and
\beq
T_{\alpha\alpha_1}(\varepsilon,\varepsilon_1) =
\gamma_e(\varepsilon)\gamma_c(\varepsilon_1)
K_{\alpha\alpha_1}(\varepsilon,\varepsilon_1) \, .
\eeq
If there is no electron-LO phonon scattering,
$K_{\alpha\alpha_1}(\varepsilon,\varepsilon_1)$ can be readily derived
from the electronic Hamiltonian ${\cal H}_{\text e}$ as
\beq
K_{\alpha\alpha_1}(\varepsilon,\varepsilon_1) =
\frac{\delta(\varepsilon -\varepsilon_1) \delta_{\alpha
\alpha_1}}{(\varepsilon -E_{\alpha})^2+(\gamma/2)^2} \, .
\eeq

The total transparency is defined as
\beq\label{transp}
T_\alpha(\varepsilon) = \int \sum_{\alpha_{1}}
T_{\alpha\alpha_1}(\varepsilon,\varepsilon_1)\, d\varepsilon_1 \, ,
\eeq
with the corresponding correlation function
\beq\label{kav}
K_\alpha(\varepsilon) = \int \sum_{\alpha_{1}}
K_{\alpha\alpha_1}(\varepsilon,\varepsilon_1)\, d\varepsilon_1 \, .
\eeq
In the wide band approximation, where the energy dependence of
$\gamma_e$ and $\gamma_c$ can be neglected, $T_\alpha(\varepsilon)$ and
$K_\alpha(\varepsilon)$ are simply connected as
\beq\label{tggk}
T_\alpha (\epsilon) = \gamma_e\gamma_c K_\alpha (\varepsilon ) \, .
\eeq
Therefore, our task is to investigate the correlation function
$K_\alpha(\varepsilon)$.

In terms of diagrams, $K_\alpha(\varepsilon)$ is represented in Fig.~4.
In this figure the arrowed double line towards right (or left) is the
dressed one-electron retarded (or advanced) Green's function
$G_R(\alpha,\varepsilon)$
[or $G_A(\alpha,\varepsilon -\Omega)|_{\Omega\rightarrow 0}$] of the
entire system including the electron-phonon interaction, where the
horizontal arrowed single lines are the corresponding bare one-electron
Green's functions $G_R^{(0)}(\alpha,\varepsilon)$ and
$G_A^{(0)}(\alpha,\varepsilon -\Omega)|_{\Omega\rightarrow 0}$
of the electronic system without the electron-phonon interaction. If an
arrowed single line is tilted away from the horizontal direction, the
variables in the corresponding Green's function should be summed up. All
retarded (or advanced) Green's function are analytically continued into
the upper (or lower) half plane of the complex variable $\varepsilon$.
Based on this diagram, $K_\alpha(\varepsilon)$ consists of two terms
\beq\label{ka}
K_\alpha (\varepsilon) = K_\alpha^0 (\varepsilon) +
K_\alpha^v (\varepsilon) \, .
\eeq
The first term is just a product of the two analytically continued
dressed one-electron Green's functions
\beq\label{ka0}
K_\alpha^0 (\varepsilon) = |G_R (\alpha ,\varepsilon)|^2 \, .
\eeq
The second term is the vertex part, which can be expressed as
\beq\label{vert}
K_\alpha^v (\varepsilon) = |G_R (\alpha, \varepsilon)|^2
{\cal K}(\alpha ,\varepsilon) \, .
\eeq

Let us first consider the Green's function $G_R (\alpha, \varepsilon)$.
To study the effect of electron-LO phonon interaction on the Green's
function, we will use the Matsubara technique\cite{mahan,abrikos}. From
the Dyson equation for the Matsubara Green's function
\begin{eqnarray}
{\cal G}(\alpha ,\varepsilon)& =& {{\cal G}^{(0)}(\alpha ,\varepsilon)
\over 1 - {\cal G}^{(0)}(\alpha ,\varepsilon)
\Sigma (\alpha ,\varepsilon)}
\nonumber \\
&=& {1 \over \varepsilon - E_{\alpha}
+ \frac{i}{2}\delta - \Sigma (\alpha ,\varepsilon)} \, ,
\end{eqnarray}
where $\delta\equiv\gamma\text{sign}(\text{Im}[\varepsilon ])$, and
\beq\label{gf1}
{\cal G}^{(0)}(\alpha,\varepsilon) = \frac{1}{\varepsilon -
E_\alpha+ i\delta /2}
\eeq
with discrete Fermion frequencies $\varepsilon =i\pi(2n+1)k_{\text B}T$,
it is clear that we need to calculate the self-energy
$\Sigma (\alpha ,\varepsilon)$. While the self-energy will be analyzed
in details later to all orders of electron-phonon interaction, to the
lowest order shown in Fig.~5a, we have
\begin{eqnarray} \label{se}
\Sigma(\alpha,\varepsilon) &=& -k_{\text
B}T\sum_{\varepsilon_1,\alpha_1,{\bf q}}
{\cal G}^{(0)}(\alpha_1,\varepsilon_1)
{\cal D}^{(0)}({\bf q},\varepsilon - \varepsilon_1) \nonumber \\
& & \times \mid M_{\alpha\alpha_1}({\bf q})\mid^2 \, ,
\end{eqnarray}
where
\beq
{\cal D}^{(0)}({\bf q},\omega) = \frac{2\hbar\omega_0(q)}{\omega^2 -
\hbar^2\omega_0^2(q)}
\eeq
with discrete Boson frequencies $\omega =i\pi2nk_{\text B}T$.

To the same order as the above self-energy, the vertex correction
${\cal K}(\alpha,\varepsilon)$ in (\ref{vert}), with the corresponding
diagram Fig.~5b, can be expressed as
\begin{eqnarray}\label{vc}
& & {\cal K}(\alpha,\varepsilon) =
2\pi k_{\text B}T \sum_{\alpha_1,\varepsilon_1,{\vec q}}
\mid M_{\alpha\alpha_1}({\bf q})\mid^2 \nonumber \\
& & \times {\cal D}^{(0)}(\varepsilon -\varepsilon_1)
{\cal G}^{(0)}(\alpha_1,\varepsilon_1)
{\cal G}^{(0)}(\alpha_1,\varepsilon_1 -\Omega ) \, .
\end{eqnarray}
The factor $2\pi k_{\text B}T$ in front of this expression is due to
the fact that
the $\varepsilon_1$ integration in (\ref{transp}) has been replaced by
a summation over the discrete Matsubara energies. Comparing this
expression to (\ref{se}) for self-energy, we expect a relation between
${\cal K}(\alpha,\varepsilon)$ and $\Sigma(\alpha,\varepsilon)$. This
will be clarified in the next section.

Now we have set up the scheme to calculate the transmission probability
with (\ref{tggk})-(\ref{vert}) and (\ref{vc}). Knowing the transmission
probability, the tunneling current through a DBRTS can be calculated
from \cite{wilk}
\begin{eqnarray}\label{cur1}
J(V) &=& \frac{e}{\pi\hbar}
\int \sum_{\alpha} T_\alpha(\varepsilon) \nonumber \\
& & \times \left[ f_e(\varepsilon -eV) -f_c(\varepsilon) \right]
d\varepsilon \, ,
\end{eqnarray}
where
\[
f_j(\varepsilon) = \left[\exp \left( \frac{\varepsilon
- \epsilon_{j,{\text F}}}{k_{\text B}T} \right) + 1 \right]^{-1}
\]
is the Fermi distribution function in the emitter (or collector) for
$j=e$ (or $j=c$) with the Fermi energy $\epsilon_{e,{\text F}}$ (or
$\epsilon_{c,{\text F}}$). For realistic samples of DBRTS, resonant
tunneling occurs at a bias $eV$ much larger than the thermal energy
$k_{\text B}T$. Under this condition, we have
$f_c(\varepsilon)\simeq 0$. Hence, the {\em differential} conductance
$G(eV)\equiv (\partial J/\partial V)$ has the simple form
\beq\label{dc}
G(eV) = \frac{e^2}{\pi\hbar} \int_{-\infty}^{\infty}
\sum_{\alpha} T_\alpha (\varepsilon)
\left(-\frac{\partial f_e(\varepsilon -eV)}{\partial
\varepsilon} \right) \, d\varepsilon \, .
\eeq

\section{Self-energy and Vertex Correction}

We will first calculate the lower order self-energy (\ref{se}) and
vertex correction (\ref{vc}), and then generalize the results to the
sum over terms of all higher orders.

Let us start with the self-energy. The analytical continuation of the
self-energy (\ref{se}) has been performed by Abrikosov et.al.
\cite{abrikos} in details. The key step is to treat $\varepsilon_1$ as
a continuous complex variable and rewrite the summation over
$\varepsilon_1$ into an integral over the path $C$ as indicated by the
solid arrowed curves in Fig.~6a. In this contour integral, the
contribution comes from the horizontal path Im$[\varepsilon_1]=0$ and
Im$[\varepsilon_1]=\varepsilon$. The resulting self-energy is then
analytically continued into the upper half plane of the complex variable
$\varepsilon$. We then obtain
\begin{eqnarray}\label{sefin}
& & \Sigma_R(\alpha,\varepsilon) = -\left\{
\sum_{\alpha_1,\bf q}\right.
\mid M_{\alpha\alpha_1}({\bf q})\mid^2
\int {d\varepsilon_1 \over 2\pi} \nonumber \\
&\times & \left.
\text{Im}\left[ G^{(0)}_R(\alpha_1,\varepsilon_1) \right]
D^{(0)}_R({\bf q},\varepsilon -\varepsilon_1)
\tanh \left( {\varepsilon_1 \over 2k_{\text B}T}
\right) \right\} \nonumber \\
&+& G^{(0)}_R(\alpha_1,\varepsilon -\varepsilon_1)
\text{Im}\left[ D^{(0)}_R({\bf q},\varepsilon_1) \right]
\coth \left( {\varepsilon_1 \over 2k_{\text B}T } \right) \, ,
\end{eqnarray}
where $D^{(0)}_R({\bf q},\varepsilon)$ is a bare phonon Green's
function. The contribution from the pole at $\varepsilon_1=i\varepsilon$
in Fig.~6a is included in the last term at the right hand side of
(\ref{sefin}). Using the spectral representation
\beq
G_R^{(0)}(\alpha,\varepsilon) = {1 \over \pi} \left[ \int d\varepsilon_1
{\text{Im} \left[ G_R^{(0)}(\alpha,\varepsilon_1) \right] \over
\varepsilon_1 - \varepsilon -i\delta}
\right]_{\delta\rightarrow +0}
\eeq
and a similar form for $D_R^{(0)}({\bf q},\varepsilon)$, we arrive at
the final form for the lower order self-energy
\begin{eqnarray}\label{sigma}
\Sigma_R(\alpha,\varepsilon) &=& -{1 \over \pi^2} \sum_{\alpha_1,\bf q}
\mid M_{\alpha\alpha_1}({\bf q})\mid^2
\int d\omega \int d\varepsilon_1 \nonumber \\
&\times& { \text{Im} \left[ D_R^{(0)}({\bf q},\omega) \right]
\text{Im} \left[ G_R^{(0)}(\alpha_1,\varepsilon_1) \right] \over
\omega + \varepsilon_1 - \varepsilon - i\delta} \nonumber\\
&\times& \left[ N(\omega)+1-f(\varepsilon_1) \right] \, ,
\end{eqnarray}
where $N(\omega)$ is the Planck distribution function.

To further analyze the above expression, we will adopt a dispersionless
bulk LO-phonon spectrum $\omega_0({\bf q})\simeq\omega_0$, which is the
commonly used approximation. In this case the phonon Green's functions
becomes ${\bf q}$-independent
$D_R^{(0)}({\bf q},\omega)\simeq D_R^{(0)}(\omega)$. Since the system
is spatially homogeneous in the $(x,y)$-plane,the electronic Green's
function does not depend on $k_{1y}$, so
$G_R^{(0)}(\alpha_1,\varepsilon_1)=G_R^{(0)}(n_1,\varepsilon_1)$.
Consequently, in (\ref{sigma}) the summation over $({\bf q},k_{1y})$
acts on the matrix elements $M_{\alpha\alpha_1}({\bf q})$ alone, and
the result derived in Appendix~\ref{app1} is
\begin{eqnarray}
&&\sum_{k_{1y},{\bf q}}\mid M_{\alpha\alpha_1}({\bf q})\mid ^2 =
{M^2 \over 4\pi^2 l} A_{nn_1} \, , \label{msum} \\
A_{nn_1} &\equiv& \sqrt{2}\pi\frac{p_2!}{p_1!}
\int_0^{\infty} d\zeta\, \zeta^{2(p_1-p_2)}e^{-\zeta^2} \nonumber \\
& & \times \left[ L_{p_2}^{p_1-p_2}(\zeta^2) \right]^2 \, . \label{me}
\end{eqnarray}
Here $L_{p_2}^{p_1-p_2}(\zeta^2)$ is the Laguerre polynomial, and $p_1$
(or $p_2$) is the larger (or smaller) one of $(n,n_1)$.

What left to be done in (\ref{sigma}) is the two integrations. The
integration over $\omega$ is trivial because of the dispersionless
phonon spectrum $\text{Im}[D_R^{(0)}(\omega)]=-\pi\left[
\delta (\omega-\omega_0)-\delta (\omega+\omega_0) \right]$.
Since in realistic DBRTS samples the $\gamma$ is less than 0.5 $meV$
which is much less than the thermal energy $k_{\text B}T$ in the
problem of our interest here, the integration over $\varepsilon_1$ can
be carried out with the residual method. Together with the equality
$N(-\omega)=-1- N(\omega)$, we obtain the simple form
\begin{eqnarray}\label{lo1}
\Sigma_R(\alpha,\varepsilon) &=& \frac{M^2}{4\pi^2l}
\sum_{n_1}A_{nn_1} \left[ {N(\hbar\omega_0)+f(E_{n_1}) \over
\varepsilon -E_{n_1}+\hbar\omega_0+i\gamma/2} \right. \nonumber \\
& & + \left. {N(\hbar\omega_0)+1-f(E_{n_1}) \over
\varepsilon -E_{n_1}-\hbar\omega_0+i\gamma/2} \right] \, .
\end{eqnarray}
The important quantity is the imaginary part
\begin{eqnarray}\label{s}
\text{Im}\left[ \Sigma_R(\alpha,\varepsilon) \right] &=&
- \frac{M^2}{4\pi^2l} \sum_{n_1} A_{nn_1} \nonumber \\
& & \frac{\gamma}{2} \left[ {N(\hbar\omega_0)+f(E_{n_1}) \over
(\varepsilon -E_{n_1}+\hbar\omega_0)^2 +
(\gamma^2/4)} \right. \nonumber \\
& & \left. + {N(\hbar\omega_0)+1-f(E_{n_1}) \over
(\varepsilon -E_{n_1}-\hbar\omega_0)^2 + (\gamma^2/4)} \right]
\end{eqnarray}
which will be discussed later.

Now we come to the vertex correction ${\cal K}(\alpha,\varepsilon )$
in (\ref{vc}). The procedure of analytical continuation is similar to
that for the self-energy $\Sigma_R(\alpha,\varepsilon)$, but using
the contour of integration in Fig.~6b. After all algebraic
manipulations, we obtain
\begin{eqnarray}\label{K}
{\cal K}(\alpha,\varepsilon) &=& 2\pi\sum_{\alpha_1,{\bf q}}
\mid M_{\alpha\alpha_1}({\bf q})\mid^2
\int {d\varepsilon_1 \over 2\pi}
\left[ {\cal P}\tanh\left( {\varepsilon_1 \over 2k_{\text B}T} \right)
\right. \nonumber \\
&&\left.
+ {\cal Q}\coth\left( {\varepsilon_1 \over 2k_{\text B}T} \right)
\right] \, ,
\end{eqnarray}
where
\begin{eqnarray}
{\cal P} &\equiv& D_A^{(0)}(\varepsilon -\varepsilon_1-\Omega )\,
G_R^{(0)}(n_1,\varepsilon_1+\Omega)\,
\text{Im}[G_R^{(0)}(n_1,\varepsilon_1)] \nonumber \\
&& + D_R^{(0)}(\varepsilon -\varepsilon_1-\Omega)\,
G_A^{(0)}(n_1,\varepsilon_1)\,
\text{Im}[G_R^{(0)}(n_1,\varepsilon_1+\Omega)] \nonumber\\
{\cal Q} &\equiv & \text{Im}[D_R^{(0)}(\varepsilon_1)] \nonumber \\
&& \times G_R^{(0)}(n_1,\varepsilon -\varepsilon_1)\,
G_A^{(0)}(n_1,\varepsilon -\varepsilon_1-\Omega) \, .
\end{eqnarray}

At this stage, one can set $\Omega \rightarrow 0$. Following the same
argument which leads (\ref{sigma}) to (\ref{lo1}), the vertex
correction is derived as
\begin{eqnarray}
{\cal K}(\alpha,\varepsilon) &=& \frac{M^2}{2\pi l}
\sum_{n_1} A_{n n_1} \left[
{ N(\hbar\omega_0)+f(\varepsilon +\hbar\omega_0) \over
(\varepsilon -E_{n_1}+\hbar\omega_0)^2 +
(\gamma^2/4)} \right. \nonumber \\
&+& \left. { N(\hbar\omega_0)+1-f(\varepsilon -\hbar\omega_0) \over
(\varepsilon -E_{n_1}-\hbar\omega_0)^2 +(\gamma^2/4)} \right] \, .
\end{eqnarray}
Comparing this expression with (\ref{s}), we see that at the lowest
order of electron-phonon interaction, the self-energy and the vertex
correction have similar structure. In particular, for the situation
very close to resonance $\varepsilon\pm\omega_0\approx E_{n_1}$,
which is the main interest of the present work, we establish the
following relation
\begin{equation}\label{kf}
{\cal K}(\alpha,\varepsilon) = { 4\pi \mid \text{Im}\left[
\Sigma_R(\alpha,\varepsilon) \right] \mid \over \gamma }
\end{equation}
between the self-energy and the vertex correction. It is important to
point out that the above relation (\ref{kf}) holds
up to any order of electron-phonon coupling, provided that the
phonon spectrum is assumed dispersionless \cite{gur,lev1,lev2}.
Consequently, from now on we only need to analyze the single electron
self-energy.

All above results were derived with the only approximation that the
weak dispersion of the LO phonon spectrum can be ignored. If the
tunneling is not resonant, these results are accurate enough to be
compared with experiments at all temperatures. They are also reliable
if at low temperature the thermal energy is much smaller than the
LO phonon energy. However, since we are interested in resonant
magnetotunneling at finite temperatures, higher order electron-phonon
interaction must be included in the self-energy and the vertex
correction. This problem is exactly solvable if only two Landau levels
$n=0$ and $n=1$ in the well dominate the tunneling process. As
described at the end of Sec.~II for our model, this two-level system
can be realized by fixing the concentration of impurities in the
emitter. For the phonon absorption resonant magnetotunneling process,
the tunneling electrons are emitted from the Landau level $n_e=0$ in
the emitter. On the other hand, for the phonon emission resonant
magnetotunneling process, the emitting Landau level is $n_e=1$. From
now on we will restrict ourselves to such two-level system.

The procedure of calculating the contribution to self-energy by any
order of electron-phonon interaction is the same as that by the lowest
order. Analysis of analytical continuation leads to the following rule
of constructing diagrams. We introduce two kinds of phonon lines: a
dotted line describes a absorption process with a Bose factor
$N(\hbar\omega_0)$, while a dashed one corresponds to a phonon emission
with a factor $N(\hbar\omega_0)+1$. Then, each dotted line should be
followed by a dashed line, and vice versa. Consequently, all even order
contributions vanish, and only the diagrams with odd number of phonon
lines survive. As an example, the third order diagram is shown in
Fig.~7. Based on the diagrammatic analysis, the $\nu$th order
contribution contains $2\nu$ electron-phonon vertices, $\nu$ factors
$g^2A_{01}$ with $g^2(B)\equiv M^2/4\pi^2l$, $\nu$ phonon factors
$N(\omega_0)+1/2\mp 1/2$, and $2\nu -1$ electron Green's functions.
Therefore, the final form of each order's contribution is very similar
to (\ref{lo1}), except that there is only one term in the summation of
$n_1$, and $A_{01}\simeq 1$.

We will not present here the detailed calculation, but will only show
the final results. Since our model neglects the electron-electron
interaction in the well, it is equivalent to the case that in the well
the occupation probability of each electronic eigenstate is negligible.
Then, including all orders in $g$, the electron self-energy is
obtained as
\begin{eqnarray}\label{self1}
& & \Sigma_R(\alpha,\varepsilon) = \nonumber \\
& & \sum_{s=0}^{\infty} \left[
{ g^{2(2s+1)}\, N^{s+1}(N+1)^s \over
(\varepsilon -E_{n_1}+\hbar\omega_0+i\gamma/2)^{2s+1}\,
(\varepsilon -E_{n}+i\gamma/2)^{2s} } \right. \nonumber \\
& & + \left. { g^{2(2s+1)}\, N^s
(N+1)^{s+1} \over
(\varepsilon -E_{n_1}-\hbar\omega_0+i\gamma/2)^{2s+1}\,
(\varepsilon -E_{n}+i\gamma/2)^{2s} } \right] \,
\end{eqnarray}
where $N \equiv N(\hbar \omega_0)$ is the Planck function.
When $\varepsilon\simeq E_0$ and so $E_{n_1}=E_1$, the first term in
(\ref{self1}) gives the phonon absorption resonant magnetotunneling.
On the other hand, when $\varepsilon\simeq E_1$ and so $E_{n_1}=E_0$,
the phonon emission resonant magnetotunneling is represented by the
second term in (\ref{self1}).

\section{Self-consistent renormalization}

Studies on resonance effects at very low temperatures in both free
electron gas \cite{lev1,lev2} and in DBRTS\cite{boeb,zcg} indicate
that resonant states are strongly coupled, and the perturbation
expansion should be renormalized. A so-called self-consistent
renormalization can be used, which replaces all bare electron Green's
functions in the self-energy formula with the corresponding dressed
ones, resulting in self-energy equations.

Let us first renormalize the first order self-energy to review the
phonon emission process in which the tunneling electron starts from
the Landau level $n_e=1$ in the emitter. The self-consistent equation
is simply
\beq\label{sc1}
\Sigma_R(\alpha,\varepsilon) = \frac{g^2}{\varepsilon -
E_0-\hbar\omega_0+\Sigma_R(\alpha,\varepsilon)+i\gamma/2} \, .
\eeq
Very near to the resonant $\varepsilon -E_0-\hbar\omega_0=0$ and
$\hbar\omega_0=\hbar\omega_c$, this self-consistent equation has the
solution
\beq
\Sigma_R(\alpha,\varepsilon) \simeq
(E_0+\hbar\omega_0-\varepsilon)/2  \pm g \, ,
\eeq
which will produce the Landau levels anti-crossing phenomenon
\cite{boeb,zcg}. For a 2D free electron gas. the higher-order
corrections has been analyzed in Ref.~\onlinecite{lms1}.

Now let us turn to the main theme of the present work: finite
temperature phonon assisted resonant magnetotunneling. Here both
phonon emission and phonon absorption are involved, and we must use
(\ref{self1}) as the basis for the renormalization. If we define
$\Sigma_R(n,\varepsilon)\equiv\Sigma_R(\alpha,\varepsilon)$ and
$G_R(n,\varepsilon)\equiv G_R(\alpha,\varepsilon)$ for
$\alpha=(n,k_y)$, then the self-consistent equation can be expressed
as
\widetext
\begin{eqnarray}\label{s1}
\Sigma_R(1,\varepsilon +\hbar\omega_0) &=& \sum_{s=0}^{\infty}
g^{2(2s+1)}\, N^s(\hbar\omega_0)\, [N(\hbar\omega_0)+1]^{s+1}\,
G_R^{2s+1}(0,\varepsilon)G_R^{2s}(1,\varepsilon +\hbar\omega_0) \, ,
\nonumber \\
\Sigma_R(0,\varepsilon) &=& \sum_{s=0}^{\infty}
g^{2(2s+1)}\, N^{s+1}(\hbar\omega_0)\, [N(\hbar\omega_0)+1]^s\,
G_R^{2s}(0,\varepsilon)G_R^{2s+1}(1,\varepsilon +\hbar\omega_0) \, .
\end{eqnarray}
\narrowtext
Let us define the temperature dependent coupling constant
$\mu\equiv g^2\sqrt{N(\hbar\omega_0)[N(\hbar\omega_0)+1]}$, the
dimensionless thermal factor
$\nu\equiv\sqrt{N(\hbar\omega_0)/[N(\hbar\omega_0)+1]}$, and
\beq\label{lcomp}
L(\varepsilon) \equiv
{ \mu G_R(0,\varepsilon)G_R(1,\varepsilon +\hbar\omega_0) \over
1 - \mu^2 [G_R(0,\varepsilon)G_R(1,\varepsilon +\hbar\omega_0)]^2 } \, .
\eeq
If we further define
\[
\theta (\varepsilon) \equiv { \varepsilon
- (E_0+E_1-\hbar\omega_0-i\gamma)/2 \over \sqrt{\mu} } \, ,
\]
\[
\delta \equiv { |E_0-E_1+\hbar\omega_0| \over 2\sqrt{\mu} } \, ,
\]
then, we can rewrite (\ref{s1}) as
\begin{eqnarray}\label{s3}
\Sigma_R(1,\varepsilon +\hbar\omega_0) &=&\sqrt{\mu}
[\theta(\varepsilon)+\delta]
{ L(\varepsilon)/\nu \over 1+L(\varepsilon)/\nu } \nonumber \\
\Sigma_R(0,\varepsilon) &=&\sqrt{\mu} [\theta(\varepsilon)-\delta]
{ \nu L(\varepsilon) \over 1+\nu L(\varepsilon) } \nonumber \\
G_R(1,\varepsilon +\hbar\omega_0) &=&\frac{1}{\sqrt{\mu}}
\frac{1+L(\varepsilon)/\nu}{\theta(\varepsilon)+\delta} \nonumber \\
G_R(0,\varepsilon) &=&\frac{1}{\sqrt{\mu}}
\frac{1+\nu L(\varepsilon)}{\theta(\varepsilon)-\delta} \, .
\end{eqnarray}

Substituting the last two equations of (\ref{s3}) into (\ref{lcomp}),
we obtain the equation
\beq\label{a1}
 {[1\pm\sqrt{1+4L^2}][1+L(\nu+1/\nu) + L^2]
 \over 2L }
 = \theta^2(\varepsilon) - \delta^2
\eeq
to solve for $L(\varepsilon)$. Knowing $L(\varepsilon)$, we go back to
the first two equations of (\ref{s3}) to calculate the renormalized
self-energies $\Sigma_R(1,\varepsilon +\hbar\omega_0)$ and
$\Sigma_R(0,\varepsilon)$.

\section{Results}

We will solve (\ref{a1}) for a temperature corresponding to
$\gamma\ll\sqrt{\mu}$. There are several branches of solutions, and
the proper one should be chosen according to the correct analytical
properties and asymptotic behavior of the Green's functions. In
general, the asymptotics of Green's functions at large $\varepsilon$
is $1/\varepsilon$. Consequently, the correct branch of solutions
should correspond to
$\lim_{\varepsilon\rightarrow\infty}L(\varepsilon )=0$.

Let us first check a limiting resonant tunneling case.
At $\mid |\theta(\varepsilon)|^2-\delta^2 \mid \ll 1$, the proper
solution corresponds to $|1+L(\varepsilon)/\nu| \ll 1$. Therefore, we
obtain $L(\varepsilon)\simeq -\nu$, which yield the solutions
$G_R(1,\varepsilon)\rightarrow 0$ and
$G_R(0,\varepsilon)\rightarrow
\sqrt{\mu}(1-\nu^2)/[\theta(\varepsilon)-\delta]$.
We see that in the phonon absorption process, the original elastic
resonant tunneling peak in the transparency spectrum survives, and its
position is not affected by the electron-phonon scattering. This is in
contrast to the phonon emission process, where the original elastic
resonant tunneling peak splits symmetrically into a doublet of almost
equal strength. At sufficiently high temperature
$k_{\text B}T\gg\hbar\omega_0$, $\nu\rightarrow 1$ and then both
$G_R(0,\varepsilon)$ and $G_R(0,\varepsilon)$ approach zero.

This solution of limiting case will be used as the initial input when
solving the selfconsistent equation (\ref{a1}) numerically. We will
study the case that one electron tunnels into the well from the
$n_e=0$ Landau level in the emitter, under the exact resonant
condition $\delta =0$. We also choose  $\gamma =0.11g$. Two
temperatures are used in our
calculation: $T$=174 K (corresponding to $\nu$=0,3) with
Im$[\theta(\varepsilon)]$=0.1, and $T$=74 K (corresponding to
$\nu$=0,006) with Im$[\theta(\varepsilon)]$=0.23. For a
given value of Re$[\theta(\varepsilon)]$, we solve (\ref{a1}) for both
Re$[L(\varepsilon)]$ and Im$[L(\varepsilon)]$, and then plot them in
Fig.~8. The dotted contour is for $T$=174 K with
Re$[\theta(\varepsilon)]$ varies along the contour from -4 to -0.2,
and then from 0.2 to 4. For the other temperature $T$=74 K, the
solution is the crossed curve with Re$[\theta(\varepsilon)]$ varies
from -5 to -0.4, and then from 0.4 to 5.

Based on the solution in Fig.~8, the corresponding dimensionless
transmission probability $K_0(\varepsilon)g^2$ [defined by (\ref{kav})
and (\ref{tggk})] is calculated, and the results are shown in Fig.~9a
as functions of reduced energy Re$[\theta(\varepsilon)]\sqrt{\mu}/g$. In the
absence of electron-phonon scattering, the sharp peak of transmission
probability is marked as the curve 1. When the electron-phonon
interaction is turned on, two inelastic wings are formed with
increasing strength as the temperature is raised from $T$=74 K (curve
2) to $T$=174 K (curve 3). In order to demonstrate these inelastic
wings more clearly, we use (\ref{ka}) to decompose the transmission peak
$K_0(\varepsilon)$
into  the main peak $K_0^0(\varepsilon)$ and the inelastic wings
$K_0^v(\varepsilon)$. The results are plotted in Fig.~9b where curve 1)
shows the main peak (for both temperatures), and where
curve 2) and curve 3) show respectively
the inelastic wings for $T=74K$ and $T=174K$. For both
temperatures, within the accuracy of our plotting, the broadening of
$K_0^0(\varepsilon)$ due to the virtual phonon processes can not be
detected because it can not be distinguished from the pure elastic
peak. On the other hand, the temperature dependence of the inelastic
wings is quite dramatic.

The phonon absorption process is entirely different from the phonon
emission process, which we have also calculated here and found the
separation between the two resolved sharp peaks being proportional
to $2g$. The physical origin of this difference is that the phonon
emission is a coherent process for which the electron-phonon system
remains coherent, while the phonon absorption is typically an
incoherent process since absorbed real phonons have random phases.

\section{Remarks}

As demonstrated in the previous section, the absorption of real phonons
during the resonant tunneling yields only minor corrections in the
elastic main peak of the transparency spectrum. On the other hand, the
inelastic process indeed produces two broad wings in the spectrum, the
strength of which is sensitive to the resonant condition and grows with
increasing temperature.  If the tunneling
current and so the differential conductance $G(eV)$ can be measured in
this way, we can further extract informations of the dynamical tunneling
process as follows. Since the tunneling is dominated by the transmission
probability of only one channel, $T_0(\varepsilon)$, (\ref{dc}) can be
inverted to the form
\begin{eqnarray}\label{g-tr}
T_0(\varepsilon) &=& { 2 \hbar \over e^2k_{\text B}T }
\int_{0}^{\infty} d\xi \int_{-\infty}^\infty d\mu \nonumber \\
& & \times\cos\left( \frac{\varepsilon -\epsilon_{e,{\text F}}
-\mu}{k_{\text B}T}\xi \right)
\frac{\sinh \xi}{\xi} G(\mu) \, .
\end{eqnarray}
Substituting the measured $G(eV)$ into the integrand and performing a
numerical integration, the transparency spectrum $T_0(\varepsilon)$
can then be {\em measured}.

The self-energy (\ref{s}) has a similar form as that obtained by
Fertig et.al. \cite{sarma} for a resonant tunneling system interacting
with impurities under an applied magnetic field. While in both cases
the magnetic field broadens the width of the resonant tunneling peak
because of the violation of selection rules for the parallel momentum,
there are two important differences. In our case the broadening reaches
its maximum value when $\varepsilon =E_{n_1}\pm\hbar\omega_0$.
Furthermore, the self-energy due to the electron-phonon interaction is
proportional to $B^{1/2}$, but for the electron-impurity scattering
it is proportional to $B$.

The present work is partially supported by the Norwegian Research Council,
Grant No. 100267/410.

\appendix

\section{Summation of electron-LO phonon matrix elements}
\label{app1}

We will calculate the $(k_{1y},{\bf q})$ summation of the electron-LO
phonon matrix elements
\begin{eqnarray}
& &\sum_{k_{1y},{\bf q}}\mid M_{\alpha\alpha_1}({\bf q})\mid^2
= \sum_{k_{1y},{\bf q}} {M^2 \over V_0q^2}
\int dx\,dy\,dz\,dx^\prime\,dy^\prime\,dz^\prime \nonumber \\
& & \times e^{-{\bf qr}}e^{{\bf qr}^\prime}
\phi_{\alpha_1}^*({\bf r}) \phi_\alpha({\bf r})
\phi_\alpha^*({\bf r^\prime}) \phi_{\alpha_1}({\bf r^\prime}) \, .
\end{eqnarray}
First, we transform the summation into integrations. Since the Green's
functions are independent of $k_y$, we can conveniently choose $k_y =0$.
Therefore, we have
\begin{eqnarray}
& & \sum_{k_{1y},{\bf q}}\mid M_{\alpha\alpha_1}({\bf q})\mid^2
= \int{V_0 d^3{\bf q} \over (2\pi)^3} \int{l\, dk_{y1} \over 2\pi}
\mid M_{\alpha\alpha_1}({\bf q})\mid^2 \nonumber \\
& & = \int{d^3{\bf q} \over (2\pi)^3}{M^2 l \over q^2}
\int dx\, dx^\prime\, dz\, dz^\prime \nonumber \\
& & \times \chi^*(z)\chi(z)\chi^*(z^\prime)\chi(z^\prime)
e^{ilq_x(x-x^\prime)} e^{idq_z(z-z^\prime)} \nonumber \\
& & \times \varphi^*_{n_1}(x-l^2q_y) \varphi_{n_1}(x^\prime -l^2q_y)
\varphi^*_n(x^\prime) \varphi_n(x) .
\end{eqnarray}
The ${\bf q}$ integration can now be performed to give
\begin{eqnarray}
& & \int {dq_x\, dq_z \over (2\pi)^2 }
{\exp \left[ iq_x (x-x^\prime ) + iq_z (z-z^\prime) \right]
\over q_z^2+q_x^2+q_y^2} \nonumber \\
& & = {1 \over 2\pi} K_0 \left[ \mid q_y\mid
\sqrt{(z-z^\prime )^2 + (x-x^\prime)^2} \right] \, ,
\end{eqnarray}
where $K_0$ is the McDonald function.

For realistic DBRTS samples, the width $d$ of the quantum well is much
smaller than
the magnetic length $l$. In this case the difference $z-z^\prime$ in
the argument of the McDonald function can be ignored, and the
integration over $z$ and $z^\prime$ is equal to 1 because of the
normalization condition. By introducing two new dimensionless variables
$\xi\equiv q_yl$, $\eta\equiv (x-x^\prime)/l$, and $\zeta \equiv x/l$
we arrive at the final expression
\beq\label{me1}
\sum_{k_{1y},{\bf q}} \mid M_{\alpha\alpha_1}({\bf q}) \mid^2
= {M^2 \over (2\pi)^2 l } A_{nn_1} \, ,
\eeq
where
\begin{eqnarray}
A_{nn_1} & \equiv & \int d\zeta \, d\xi\, d\eta\,
K_0\left( \mid \eta\xi \mid \right) \nonumber \\
&\times& \tilde{\varphi}_{n_1}^*\left( \zeta- {\xi - \eta \over 2} \right)
\tilde{\varphi}_{n_1}\left( \zeta- {\xi +\eta \over 2} \right) \nonumber \\
&\times& \tilde{\varphi}_n^*\left( \zeta + {\eta-\xi \over 2} \right)
\tilde{\varphi}_n\left( \zeta +{\eta +\xi \over 2} \right) \, .
\end{eqnarray}
Substituting the harmonic oscillator wave functions
\beq
\tilde{\varphi}_n(\zeta) = \left( {1 \over \sqrt{\pi}2^n n!} \right)^{1/2}
e^{-\zeta^2/2} H_n (\zeta)
\eeq
into the above integrand, where $H_n(\zeta)$ is the Hermite polynomial,
we obtain the results given by (\ref{me}).

\begin{figure}
\caption{A schematic illustration of a double-barrier resonant tunneling
structure (DBRTS).}
\label{fig1}
\end{figure}

\begin{figure}
\caption{Diagrams illustrate phonon-assisted resonant magnetotunneling
as the magnetic field is tuned into the resonant condition
where the Landau level separation $\hbar\omega_c$ equals the
optical phonon energy $\hbar\omega_0$:
A) A phonon {\em emission} process leading to two coherent
final states ($E0$ and $E1$), and b) a phonon {\em absorption} process
leading to two final states ($A0$ and $A1$).}
\label{fig2}
\end{figure}

\begin{figure}
\caption{Diagrammatic representation for the two-particle transmission
Green's function $K_{\alpha\alpha_1}(\varepsilon,\varepsilon_1)$.}
\label{fig3}
\end{figure}

\begin{figure}
\caption{
Diagrammatic representation for the total transmission
$K_\alpha (\varepsilon)$ given by Eqs.~(\protect{\ref{ka}})-
(\protect{\ref{vert}}).
}
\label{fig4}
\end{figure}

\begin{figure}
\caption{Lowest order diagrams for (a) the self-energy
$\Sigma (\alpha,\varepsilon)$, and (b) the vertex Green's function
${\cal K}(\alpha,\varepsilon)$.}
\label{fig5}
\end{figure}

\begin{figure}
\caption{Contours of integration for (a) the self-energy
$\Sigma (\alpha,\varepsilon)$, and (b) the vertex Green's function
${\cal K}(\alpha,\varepsilon)$.}
\label{fig6}
\end{figure}

\begin{figure}
\caption{Resonant third order diagram for the self-energy. A dotted
line denotes a phonon absorption process, while  a dashed line denotes
an emission process. The electron lines are marked by their Landau
level numbers.}
\label{fig7}
\end{figure}

\begin{figure}
\caption{Solution Re$[L(\omega)]$ and Im$[L(\omega)]$ of
Eq.~(\protect{\ref{a1}})
for $\delta =0$ and $\gamma =0.11g$. Dotted contour is for temperature
$T$=174 K, and crossed contour is for $T$=74 K.}
\label{fig8}
\end{figure}

\begin{figure}
\caption{Phonon absorption resonant magnetotunneling spectrum
corresponding to the solution in Fig.~8. A) The dimensionless
transmission probability $K_0(\varepsilon)g^2$ is plotted
as function of the reduced energy
$\protect{\text Re\protect}\left[\theta(\varepsilon)\right]
\protect\sqrt{\mu} /g$.
Curve 1) shows the transmission probability in the absence
of electron-phonon scattering, while curve 2) and curve 3) show
respectively the results at $T=74K$ and at $T=174K$. B) Transmission
spectrum decomposed into  main peak
$K_0^0(\varepsilon)g^2$ (curve 1) and inelastic wings
$K_0^\nu(\varepsilon)g^2$ at $T=74K$ (curve 2) and $T=174k$ (curve 3).}
\label{fig9}
\end{figure}

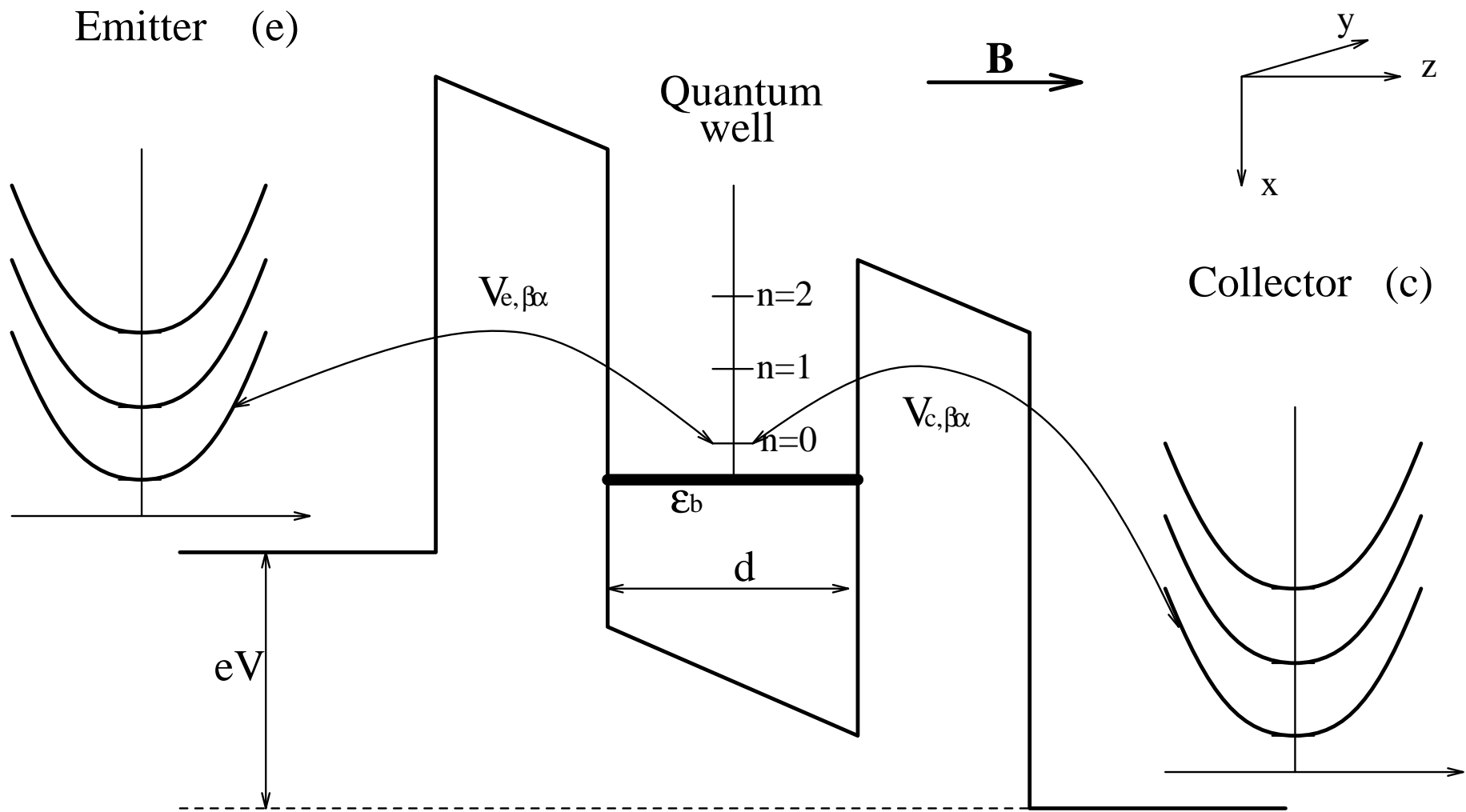
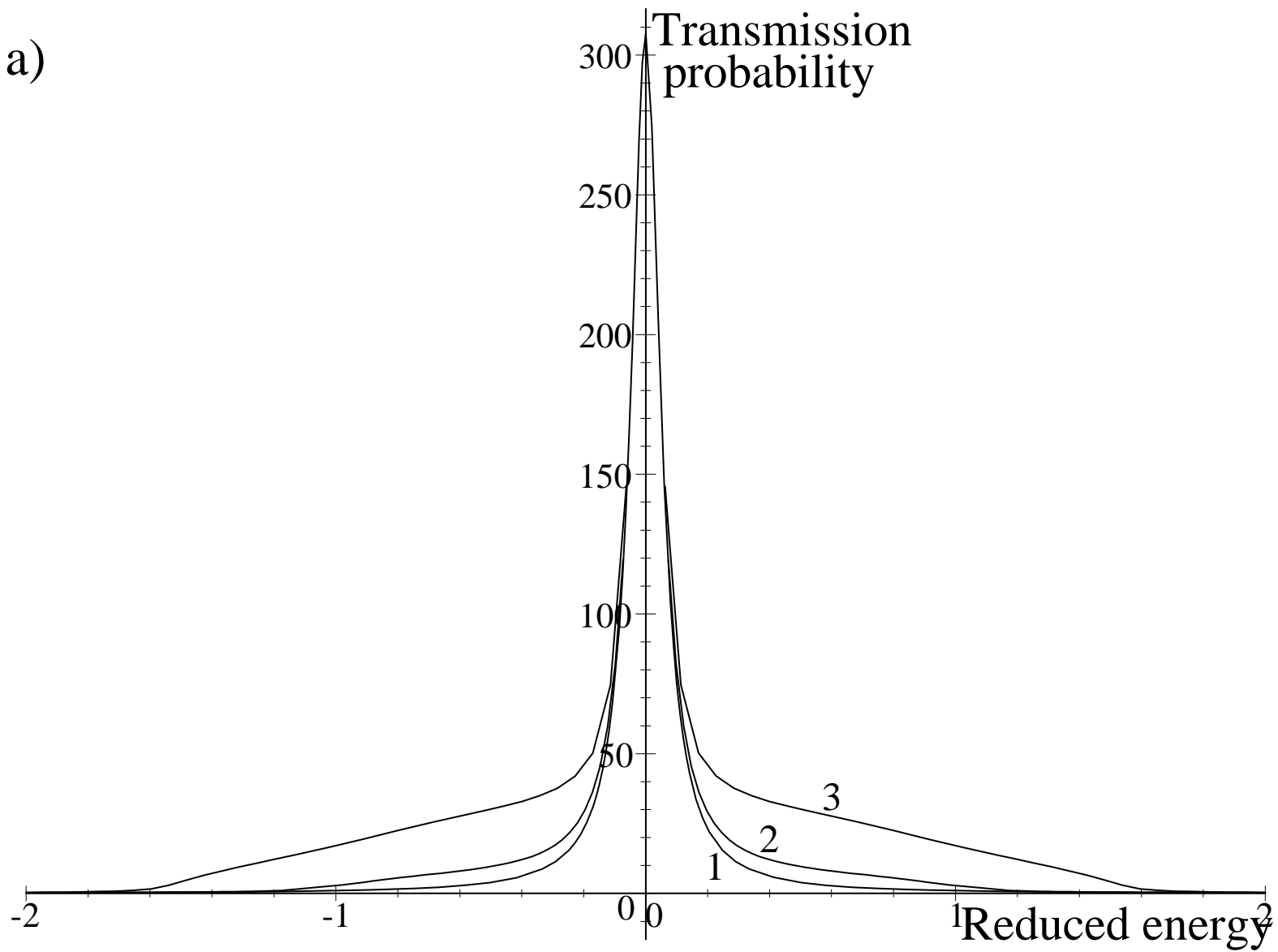
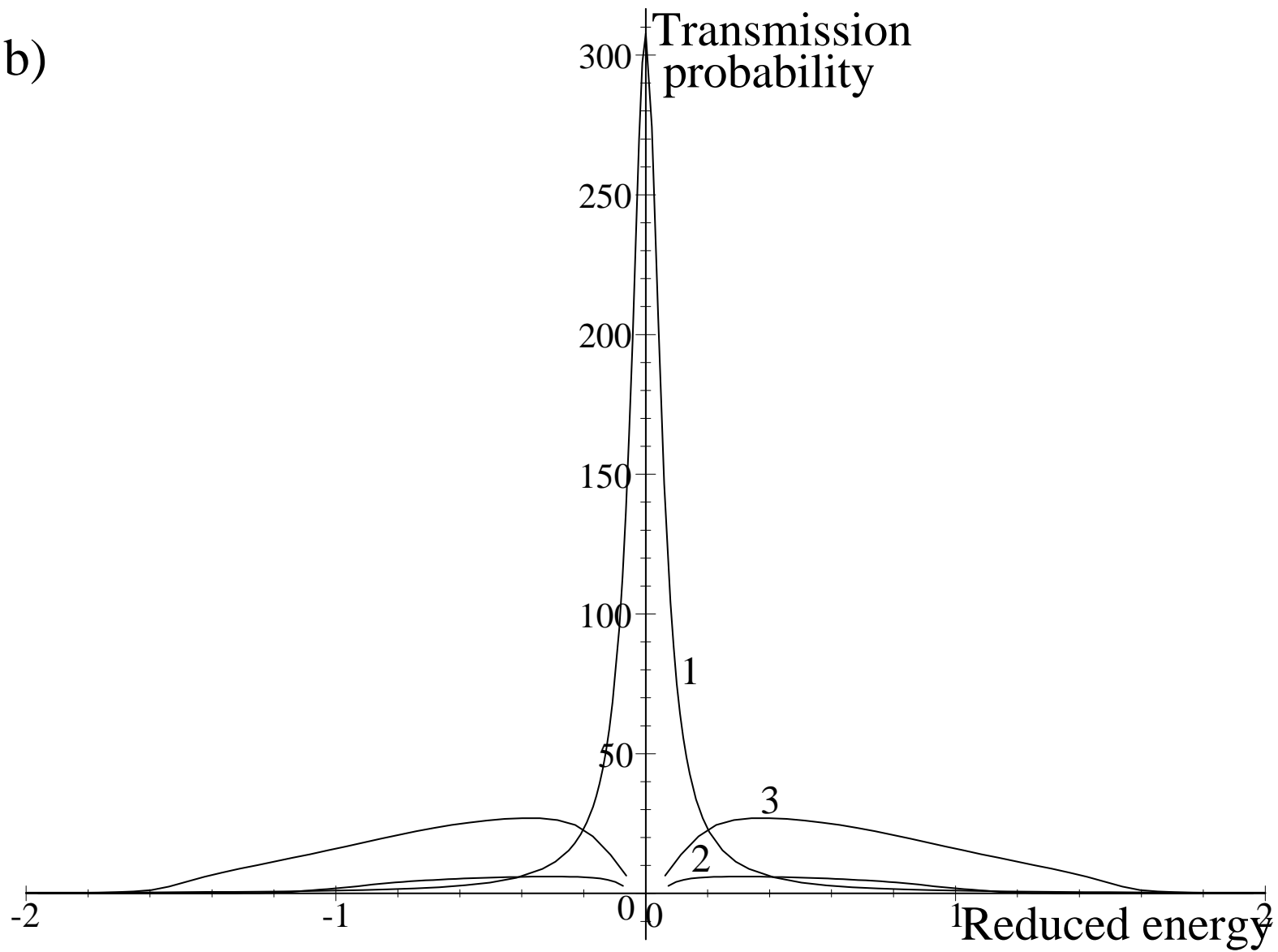
\end{document}